\documentclass[10pt,a4paper]{article}
\usepackage[utf8]{inputenc}
\usepackage[english]{babel}
\usepackage{amssymb,graphicx}
\usepackage{amsmath,amsfonts}
\usepackage{amsthm}
\usepackage{framed}
\usepackage{fullpage}
\usepackage[makeroom]{cancel}
\usepackage[export]{adjustbox} 
\usepackage{feynmf}
\usepackage{microtype}
\usepackage{hyperref}
\usepackage{mathtools}

\def\be{\begin{eqnarray}}
\def\ee{\end{eqnarray}}
\def\nn{\nonumber}

\def\p{\partial}

\newcommand{\beq}{\begin{equation}}
\newcommand{\eeq}{\end{equation}}
\newcommand{\beqa}{\begin{eqnarray}}
\newcommand{\eeqa}{\end{eqnarray}}

\def\NS{\mathfrak{P}}

\usepackage[cmtip,all]{xy}

\newcommand{\longsquiggly}{\xymatrix{{}\ar@{~>}[r]&{}}}

\newcommand*{\Warrow}{\rotatebox[origin=c]{180}{\(\xrightarrow{\hspace*{3.5cm}}\)}}
\newcommand*{\Earrow}{\rotatebox[origin=c]{0}{\(\xrightarrow{\hspace*{3.5cm}}\)}}

\begin{document}

\title{\vspace{1.5cm}\bf
A basic triad in Macdonald theory
}

\author{
A. Mironov$^{b,c,d,}$\footnote{mironov@lpi.ru,mironov@itep.ru},
A. Morozov$^{a,c,d,}$\footnote{morozov@itep.ru},
A. Popolitov$^{a,c,d,}$\footnote{popolit@gmail.com}
}

\date{ }

\maketitle

\vspace{-6cm}

\begin{center}
MIPT/TH-28/24  \hfill {\bf to the memory of}\\
  FIAN/TD-16/24 \hfill {\bf Masatoshi Noumi}\\
ITEP/TH-36/24  \hfill \phantom{.}\\
 IITP/TH-31/24  \hfill \phantom{.}
\end{center}

\vspace{3cm}

\begin{center}
$^a$ {\small {\it MIPT, Dolgoprudny, 141701, Russia}}\\
$^b$ {\small {\it Lebedev Physics Institute, Moscow 119991, Russia}}\\
$^c$ {\small {\it NRC ``Kurchatov Institute", 123182, Moscow, Russia}}\\
$^d$ {\small {\it Institute for Information Transmission Problems, Moscow 127994, Russia}}
\end{center}

\vspace{.1cm}

\begin{abstract}
Within the context of wavefunctions of integrable many-body systems, rational multivariable
Baker-Akhiezer (BA) functions were introduced by O. Chalykh, M. Feigin and A. Veselov and, in the case of the
trigonometric Ruijsenaars-Schneider system, can be associated with a reduction of the Macdonald symmetric
polynomials at $t=q^{-m}$ with integer partition labels substituted by arbitrary complex numbers. A parallel attempt to describe wavefunctions of the bispectral trigonometric Ruijsenaars-Schneider problem was made by M. Noumi and J. Shiraishi who proposed a power series that reduces to the Macdonald polynomials at particular values of parameters. It turns out that this power series also reduces to the BA functions at $t=q^{-m}$, as we demonstrate in this letter.
This makes the Macdonald polynomials, the BA functions and the Noumi-Shiraishi (NS) series a closely tied {\it triad}
of objects, which have very different definitions, but are straightforwardly related with each other.
In particular, theory of the BA functions provides a nice system of simple linear equations, while
the NS functions provide a nice way to represent the multivariable BA function explicitly with arbitrary number of variables.
\end{abstract}

\bigskip

\newcommand\smallpar[1]{
  \noindent $\bullet$ \textbf{#1}
}

\section{Introduction}

For a long time, the Macdonald polynomials \cite{Mac} were an artful but somewhat technical construction,
not included in the standard courses of theoretical physics, either because it seemed too involved and abstract
or because for too many it looked rather peculiar and \textit{ad hoc}.
The main drawbacks were, firstly, the lack of a clear algebraic construction, which from the very beginning existed
for the Schur polynomials thanks to relation to characters of simple Lie algebras,
and, secondly, the frankly exotic type of finite-difference integrable systems for which the Macdonald polynomials were providing eigenfunctions.

Nowadays the situation is changing, after integrability was recognized as a principal feature of
non-perturbative physics (of functional integrals defined exactly rather than perturbatively) \cite{UFN23},
and after establishing a distinguished class of Ding-Iohara-Miki (DIM) algebras \cite{DIM},
which inherit the main beauties of classical and Kac-Moody (affine) algebras: their representations being associated precisely with the Macdonald polynomials. Moreover, the DIM algebra gives rise to more integrable systems \cite{MMP} than people used to consider, and the original Macdonald polynomials are not sufficient to deal with all of them, thus there is a need for further generalizations, and this is an actual task for today. Fortunately, we have almost enough knowledge to handle this problem, and it is provided by the two main ingredients: the Chalykh-Feigin-Veselov \cite{ChFV} approach \cite{Cha} to Macdonald polynomials\footnote{This approach is basically a generalization of an earlier construction \cite{CV,Ves,Cha2} of eigenfunctions of the rational and trigonometric Calogero-Sutherland systems to the Ruijsenaars case.},
and the Noumi-Shiraishi (NS) generalization \cite{NS} of them to rather involved but still conceptually simple power series.
All these pieces of theory are now merging together, and they will soon become one of the cornerstones
of tomorrow's mathematical physics.

Speaking more concretely, in this letter, we are going to discuss the triad of the NS power series, the Macdonald polynomials and the corresponding rational Baker-Akhiezer (BA) functions introduced by O. Chalykh \cite{Cha} based on the earlier work \cite{ChFV} (in fact, the BA function in the case of our interest coincides with the $\psi$-function from \cite{ES}). The NS function is a function (power series) of $N$ variables $x_i=q^{z_i}$ (we use both variables $x_i$ and $z_i$) and of $N$ variables $\lambda_i$ depending on parameters $q$ and $t$. At the very particular values of generally complex variables $\lambda_i$ associated with integer partitions, the NS function is reduced to the Macdonald polynomial, which is a symmetric polynomial in $x_i$. There is another possibility of giving rise to a polynomial from the NS function: at $t=q^{-m}$, $m\in \mathbb{Z}_{\ge 0}$, it also reduces to a polynomial at {\it any} complex $\lambda_i$, but this polynomial is no longer symmetric in $x_i$. This polynomial as we explain in this letter is {\bf nothing but the BA function}. Thus, one can make out of the NS power series either a symmetric polynomial at arbitrary $t$ but discrete values of $\lambda_i$, or a non-symmetric polynomial at $t=q^{-m}$ but at arbitrary $\lambda_i$.

Note that all three objects of the triad are defined quite differently: the NS function is given by an explicit formula \cite{NS}, the BA function is defined by linear equations \cite{Cha}, and the Macdonald polynomials being symmetric functions of $x_i$ can be defined \cite{Mac} not only in terms of $x_i$ (in many different ways), but also in terms of the power sums $p_k=\sum_i^Nx_i^k$ from orthogonality conditions. Note also that the NS function is the eigenfunction of the Macdonald-Ruijsenaars operator, and this property is definitely survives the reductions so that both the Macdonald polynomials and the BA functions are still eigenfunctions of the same operator.

Our point of interest here is the {\bf boldfaced triangle} in the following diagram:

\bigskip

\hspace{-1.4cm}
\fbox{\parbox{18.6cm}
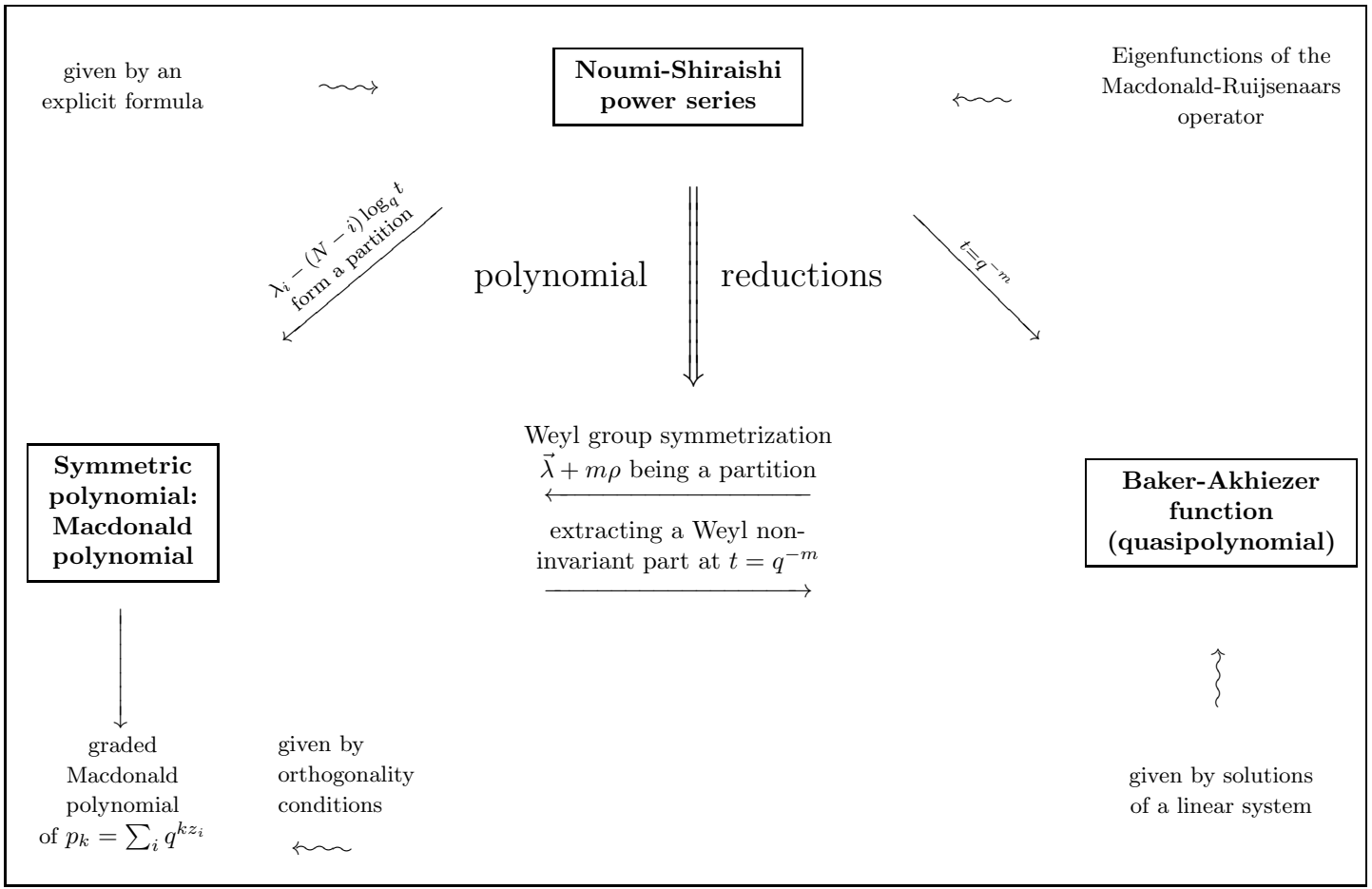
{
$$
\begin{array}{cccccr}
\begin{array}{c}\hbox{\small given by an}\cr\hbox{\small explicit formula}
\end{array}&\longsquiggly&\boxed{\begin{array}{c}
\hbox{\bf Noumi-Shiraishi}\cr
\hbox{\bf power series}
\end{array}}&{\rotatebox[origin=c]{-180}{\longsquiggly}}&\begin{array}{c}
\hbox{\small Eigenfunctions of the}\cr
\hbox{\small Macdonald-Ruijsenaars}\cr
\hbox{\small operator}
\end{array}&\\
\\
&{\rotatebox[origin=c]{40}{\(\xleftarrow{
{\footnotesize\begin{array}{c}
\lambda_i-(N-i)\log_qt\cr
\hbox{form a partition}
\end{array}}
}\)}}&\hbox{\Large polynomial {\rotatebox[origin=c]{-90}{\(\xRightarrow{{\hspace*{2.5cm}}}\)}} reductions}&{\rotatebox[origin=c]{-45}{\(\xrightarrow{\hspace*{.7cm}t=q^{-m}\hspace*{.7cm}}\)}}&\\
\\
\boxed{\begin{array}{c}\hbox{\bf Symmetric}\cr\hbox{\bf polynomial:}\cr
\hbox{\bf Macdonald}\cr\hbox{\bf polynomial}\end{array}}&&\begin{array}{c}\stackrel{\begin{array}{c}\hbox{Weyl group symmetrization}\cr \vec\lambda+m\rho\hbox{ being a partition}\end{array}}{\Warrow}\cr
\stackrel{\begin{array}{c}\hbox{extracting a Weyl non-}\cr\hbox{invariant part at }t=q^{-m}\end{array}}{\bf\Earrow}
\end{array}&
&\boxed{\begin{array}{c}\hbox{\bf Baker-Akhiezer}\cr\hbox{\bf function}\cr\hbox{\bf (quasipolynomial)}\end{array}}\\
{\rotatebox[origin=c]{-90}{\(\xrightarrow{{\hspace*{1.5cm}}}\)}}
&&&&{\rotatebox[origin=c]{90}{\longsquiggly}}\\
\begin{array}{c}
\hbox{\small graded}\cr\hbox{\small Macdonald}\cr\hbox{\small polynomial}\cr\hbox{\small of }p_k=\sum_iq^{kz_i}
\end{array}
&\begin{array}{l}\hbox{\small given by}\cr
\hbox{\small orthogonality}\cr\hbox{\small conditions}\cr{\rotatebox[origin=c]{-180}{\longsquiggly}}\end{array}
&&&\begin{array}{c}\hbox{\small given by solutions}\cr\hbox{\small  of a linear system}
\end{array}
\end{array}
$$
}}

\bigskip

\bigskip

The vertical arrow describes the two polynomial reductions of the NS functions to the Macdonald polynomials and to the BA functions. There is a connection between these two: one can obtain the Macdonald polynomial at $t=q^{-m}$ from the BA function by symmetrizing of all $x_i$ (by summing over the Weyl group). Thus Macdonald polynomials can be regarded as analytical continuation in $q^{-m}$ from the system of the BA functions \cite{Cha} (see some technical detail also in \cite{ChE,MMP1}).
This is described by the horizontal arrow.

The letter is organized as follows. In section 2, we define the NS function and describe its reduction to the Macdonald polynomial and to the BA function as well as interrelations between the latter. We also describe the principal specialization of all three, where they factorize, and explain in detail why the reductions to integer partitions and to $t=q^{-m}$ are not permutable. In section 3, we draw two more tables to describe the today situation with generalizations of the triad to the elliptic \cite{MMPZ} and twisted cases. More generalizations can be found in the Conclusion.

\paragraph{Notation.} We use the notation $M_\mu(\{x_i\};q,t)$ for the Macdonald polynomial as a symmetric polynomial of $x_i$, $i=1,\ldots,N$ labeled by the Young diagram (partition) $\mu$. Note that we always keep the number of parts of $\mu$ equal to $N$, the number of $x_i$, but some parts can be zero.

We also use the notation $\vec z$ for the vector with components $z_i$, $i=1,\ldots,N$ and similarly for $\vec\lambda$. The notation $M_{\vec\mu}(\{x_i\};q,t)$ denotes $M_\mu(\{x_i\};q,t)$ with $\mu:=[\mu_1,\ldots,\mu_N]$, $\mu_1\ge\ldots\ge\mu_N\ge 0$ and all $\mu_i$ integer.

\section{Triad}

In this section, we make detailed comments about separate items in the table.

\subsection{A basic object: NS function}

The NS function is, up to a simple pre-factor, a formal power series of $2N$ variables $q^{z_i}$ and $q^{\lambda_i}$, $i=1,\ldots,N$ generically neither symmetric, nor polynomial in $q^{z_i}$. However, it can acquire one, or both, of these properties at peculiar specializations. This power series is defined to be \cite{NS}
\be\label{NS}
\NS_{q,t}(\vec z,\vec \lambda)= q^{\vec z\cdot\vec\lambda}\cdot t^{-\vec z\cdot\vec\rho}\cdot
\left(\sum_{k_{ij}}\psi(\vec\lambda,k_{ij};q,t)\prod_{1\le i<j\le N}q^{k_{ij}(z_j-z_i)}\right)
\ee
where the sum goes over all non-negative integer $k_{ij}$ with $i< j$,
and $\vec\rho$ is the Weyl vector, i.e. $\vec\rho\cdot\vec z={1\over 2}\sum_{i=1}^N(N-2i+1)z_i$. The coefficients $\psi(\vec\lambda,k_{ij};q,t)$ are
\be
\psi(\vec\lambda,k_{ij};q,t):&=&\prod_{n=2}^N\prod_{1\le i<n}\prod_{s=0}^{k_{in}-1}{\Big(1-q^{s+1-k_{in}}t^{-1}\Big)
\over \Big(1-q^{s-k_{in}}\Big)}
{\Big(1-tq^{s+\lambda_n-\lambda_i+\sum_{a>n}(k_{ia}-k_{na})}\Big)
\over \Big(1-q^{s+1+\lambda_n-\lambda_i+\sum_{a>n}(k_{ia}-k_{na})}\Big)}
\times\\
&\times&
\prod_{n=2}^N\prod_{1\le i<j< n}\prod_{s=0}^{k_{in}-1}{\Big(1-tq^{s+\lambda_j-\lambda_i+\sum_{a>n}(k_{ia}-k_{ja})}\Big)
\Big(1-t^{-1}q^{s+1+\lambda_j-\lambda_i-k_{jn}+\sum_{a>n}(k_{ia}-k_{ja})}\Big)
\over \Big(1-q^{s+1+\lambda_j-\lambda_i+\sum_{a>n}(k_{ia}-k_{ja})}\Big)
\Big(1-q^{s+\lambda_j-\lambda_i-k_{jn}+\sum_{a>n}(k_{ia}-k_{ja})}\Big)}\nn
\label{c}
\ee
The NS function is an eigenfunction of the Macdonald-Ruijsenaars operator\cite{Mac,Rui}:
\be\label{MR}
\hat H_{MR}={\sqrt{q}\over q-1}\ \sum_{i=1}^N\prod_{j\ne i}{tq^{z_i}-q^{z_j}\over q^{z_i}-q^{z_j}}e^{\p_{z_i}}\nn\\
\hat H_{MR}\cdot\NS_{q,t}(\vec z,\vec \lambda)=\left({\sqrt{qt}\over q-1}\sum_iq^{\lambda_i}\right)\cdot\NS_{q,t}(\vec z,\vec \lambda)
\ee

\paragraph{An example of $N=2$.}

In this case, there is just one non-zero $k_{12}=k$, and, using the notation $z:=z_1-z_2$, $\lambda:=\lambda_1-\lambda_2$, one obtains
\be\label{NS2}
\NS_{q,t}(z_1,z_2,\lambda_1,\lambda_2)=q^{\lambda_1z_1+\lambda_2 z_2}\cdot t^{z\over 2}\cdot
\sum_{k=0}\psi(\lambda_1-\lambda_2,k;q,t)q^{k(z_2-z_1)}
\ee
with
\be
\psi(\lambda,k;q,t)=\prod_{s=0}^{k-1}{\Big(1-q^{s+1-k}t^{-1}\Big)
\over \Big(1-q^{s-k}\Big)}{\Big(1-tq^{s-\lambda}\Big)
\over \Big(1-q^{s+1-\lambda}\Big)}=\prod_{s=1}^{k}{\Big(1-q^{s-1}t\Big)
\over \Big(1-q^{s}\Big)}{\Big(1-t^{-1}q^{\lambda-s+1}\Big)
\over \Big(1-q^{\lambda-s}\Big)}
\label{c2}
\ee

\subsection{Reduction to Macdonald polynomials}

It is well-known that the polynomial eigenfunctions of the Macdonald-Ruijsenaars Hamiltonian (\ref{MR}) are the Macdonald polynomials \cite{Mac}. Choosing other boundary conditions, one can obtain more solutions (see, e.g., \cite{Khar}). The NS functions provides an example of solutions that are power series instead of polynomials, they are not symmetric functions, in contrast with the Macdonald polynomials, but instead they are parameterized by $N$ arbitrary complex parameters, while, in the Macdonald case, there is an integrality requirement.

However, the power series (\ref{NS}) can be made a symmetric polynomial in $q^{z_j}$ by choosing $\vec\lambda=\vec\mu+\vec\rho\log_q t$, where $\vec\mu$ has all non-negative integer components $\mu_j$, $\mu_1\ge\mu_2\ge\ldots\ge\mu_N\ge0$ at $j=1,\ldots,N$. This symmetric polynomial is nothing but the Macdonald polynomial \cite{NS}:
\be\label{NSM}
\NS_{q,t}(\vec z,\vec \mu+\vec\rho\log_q t)= M_\mu(\{q^{z_i}\};q,t)
\ee
Notice the peculiar $t$-dependent shift: it turns out to be crucial
in non-commutativity of triad specializations, see sec.\ref{sec:two-non-perm-reductions}.

This case admits a reformulation in terms of power sums $p_k:=\sum_{i=1}^Nq^{kz_i}$.

\paragraph{An example of $N=2$.}

In this case, using the notation $x_{1,2}:=q^{z_{1,2}}$, $\mu:=\mu_1-\mu_2$, one obtains from (\ref{NS2}), (\ref{c2}) and (\ref{NSM}):
\be
M_{[\mu_1,\mu_2]}(\{x_1,x_2\};q,t)= x_1^{\mu_1}x_2^{\mu_2}\cdot
\sum_{k=0}\left({x_2\over x_1}\right)^{k}\prod_{s=1}^{k}{\Big(1-q^{s-1}t\Big)
\over \Big(1-q^{s}\Big)}{\Big(1-q^{\mu-s+1}\Big)
\over \Big(1-tq^{\mu-s}\Big)}
\ee
which coincides with \cite[Eq.(5)]{MMP1}.

\subsection{Reduction to BA functions}

Instead of getting symmetric polynomials with integrality requirements for $\vec\lambda$, one can choose $t=q^{-m}$, which also gives rise to a (quasi)polynomial\footnote{We use the term quasipolynomial because of a generally non-polynomial trivial pre-factor $q^{\vec\lambda\cdot\vec z}$.} of $q^{z_i}$ though non-symmetric, but instead $\vec\lambda$ can be kept to be $N$ arbitrary complex parameters. Hence, on one hand, it does not admit a reformulation in terms of power sums $p_k:=\sum_{i=1}^Nq^{kz_i}$, but, on the other hand, this (quasi)polynomial is nothing but the BA function $\Psi_m(\vec z,\vec\lambda)$ \cite{Cha}:
\be\label{NBA}
\Psi_m(\vec z,\vec\lambda)={\cal N}_\lambda\cdot\NS_{q,q^{-m}}(\vec z,\vec \lambda)
\ee
where ${\cal N}_\lambda$ is a normalization factor chosen in such a way that the BA function is a symmetric function in $\vec z$ and $\vec\lambda$ \cite{Cha,MMP1}:
\be
{\cal N}_\lambda=\prod_{k>l}\prod_{j=1}^m\Big(q^{{\lambda_k-\lambda_l\over 2}-j}-q^{-{\lambda_k-\lambda_l\over 2}}\Big)\nn\\
\Psi_m(\vec z,\vec\lambda)=\Psi_m(\vec \lambda,\vec z)
\ee
It definitely remains to be an eigenfunction of the Macdonald-Ruijsenaars Hamiltonian.

The proof of (\ref{NBA}) is simple: both $\Psi_m(\vec z,\vec\lambda)$ and $NS_{q,q^{-m}}(\vec z,\vec \lambda)$ satisfy the same Macdonald-Ruijsenaars equation \cite{NS,Cha}, and as polynomials of $q^{z_i}$ have exactly the same terms up to coefficients. The coefficients are fixed by the Macdonald-Ruijsenaars equation unambiguously up to total normalization.

\paragraph{Macdonald polynomials from the BA functions.} One can make the Macdonald polynomial from the BA function at non-negative integer components of $\vec\lambda+m\vec\rho$ ordered in non-increasing order (and this is Chalykh's original relation between the two, see \cite[Thm.5.11]{Cha}):
\be\label{MPsi}
M_{\vec\lambda+m\vec\rho}(\{q^{z_i}\};q,q^{-m})=
{\cal N}_\lambda^{-1}\cdot\sum_{w\in W}\Psi_m(w\vec z,\vec\lambda)
\ee
Here $w$ is an element of the Weyl group of $A_{N-1}$.

Moreover, as a counterpart of the Poincare symmetry $t\to q/t$ of the NS function \cite[Theorem 6.5]{NS},
\be
\NS_{q,q/t}(\vec z,\vec \lambda)=\left({q\over t^2}\right)^{\vec z\cdot\vec\rho}\cdot\left(\prod_{1\le i<j\le N}\prod_{n=0}^\infty
{1-t^{-1}q^{n+1+z_j-z_i}\over 1-tq^{n+z_j-z_i}}\right)
\cdot\NS_{q,t}(\vec z,\vec \lambda)
\ee
one can similarly construct \cite{Cha} the Macdonald polynomial at $t=q^{m+1}$
\be\label{MPsi2}
M_{\vec\lambda-(m+1)\vec\rho}(\{q^{z_i}\};q,q^{m+1})=
{\cal N}_\lambda^{-1}\cdot
\prod_{k>l}\prod_{j=-m}^m\left(q^{j+{z_k-z_l\over 2}}-q^{z_l-z_k\over 2}\right)^{-1}\cdot\sum_{w\in W}(-1)^{|w|}\Psi_m(w \vec z,\vec\lambda)
\ee
where $|w|$ denotes the order of the Weyl group element.

In other words, while (\ref{MPsi}) implies that, at $t=q^{-m}$, the Macdonald polynomial is naturally represented by a sum of Weyl non-invariant pieces, (\ref{MPsi}) means that, at $t=q^{m+1}$, only the combination
$$
M_{\vec\lambda-(m+1)\vec\rho}(\{x_i\};q,q^{m+1})\cdot\prod_{k>l}\prod_{j=-m}^m\left(q^j\sqrt{x_k\over x_l}-\sqrt{x_l\over x_k}\right)
$$
does so. This is related with the fact that the Poincare symmetry is not that evident at the level of the Macdonald polynomials: the transition from the NS functions to the Macdonald polynomials involves the parameter $t$, which breaks down this symmetry.

\subsection{Principal specialization}

It is well-known that the Macdonald polynomials factorize under specialization $x_i=t^i$ (see, e.g., \cite{AKMM1}):
\be
M_\lambda(t^{i})=M_\lambda(t^{N-i})=t^{\sum_ii\lambda_i}\cdot
\prod_{i<j}{(q^{\lambda_i-\lambda_j}t^{j-i};q)_\infty\over (q^{\lambda_i-\lambda_j}t^{j-i+1};q)_\infty}
{(t^{j-i+1};q)_\infty\over (t^{j-i};q)_\infty}\ \stackrel{t=q^k}{=}\ t^{\sum_ii\lambda_i}\cdot
\prod_{r=0}^{k-1}\prod_{i<j}{1-q^{r+\lambda_i-\lambda_j}t^{j-i}\over 1-q^rt^{j-i}}
\ee
This factorization property is inherited from the factorization property of the NS function \cite[Theorem 6.6]{NS},
\be
\NS_{q,t}(z_i=(N-i)\log_q t,\vec \lambda)=t^{{1\over 12}N(N^2-1)\log_q t}\cdot t^{\sum_i(N-i)\lambda_i}\cdot
\prod_{i=1}^N\prod_{n=1}^\infty{1-q^nt^{-1}\over 1-q^nt^{-i}}\cdot\prod_{1\le i<j\le N}
\prod_{n=1}^\infty{1-q^{n+\lambda_j-\lambda_i}t^{-1}\over 1-q^{n+\lambda_j-\lambda_i}}
\ee
and, hence, implies the corresponding factorization of the BA function:
\be
\Psi_m(z_i=(i-N)m,\vec \lambda)={\cal N}_\lambda\cdot q^{{m^2\over 12}N(N^2-1)}\cdot q^{m\sum_i(i-N)\lambda_i}\cdot
\prod_{i=1}^N\prod_{n=1}^{m(i-1)}(1-q^{n+m})\cdot\prod_{1\le i<j\le N}
\prod_{n=1}^m{1\over 1-q^{n+\lambda_j-\lambda_i}}
\ee
Similarly,
\be
\Psi_m(z_i=-im,\vec \lambda)={\cal N}_\lambda\cdot q^{{m^2\over 12}N(N^2-1)}\cdot q^{-m\sum_ii\lambda_i}\cdot
\prod_{i=1}^N\prod_{n=1}^{m(i-1)}(1-q^{n+m})\cdot\prod_{1\le i<j\le N}
\prod_{n=1}^m{1\over q^{\lambda_i-\lambda_j}-q^n}
\ee
Note that a dependence on $\vec\lambda$ in the two last formulas is present only in the denominators of these expressions and in the simple pre-factors with opening a possibility of existing a generalization that adds analogs of
$\lambda$-parameters to the numerators, in a manner similar to (basic) hypergeometric
functions.

\subsection{On two non-permutable reductions from the NS function\label{sec:two-non-perm-reductions}}

Let us stress that, making the reduction (\ref{NSM}) at discrete values of $\vec\lambda$, one obtains a symmetric polynomial, while making the reduction (\ref{NBA}) at $t=q^{-m}$, one obtains a non-symmetric polynomial that does not become symmetric after further restricting $\vec\lambda$ according to (\ref{NSM}). In other words, the two reductions are not permutable.

Let us demonstrate how this works in the simplest case of $N=2$ (\ref{NS2}). In this case, the power series
\be
x_1^{\mu_1}x_2^{\mu_2}\cdot
\sum_{k=0}\left({x_2\over x_1}\right)^{k}\prod_{s=1}^{k}{\Big(1-q^{s-1}t\Big)
\over \Big(1-q^{s}\Big)}{\Big(1-q^{\mu-s+1}\Big)
\over \Big(1-tq^{\mu-s}\Big)}
\ee
becomes the Macdonald polynomial $M_{[\mu_1,\mu_2]}$ at integer $\mu_{1,2}$, $\mu=\mu_1-\mu_2\ge 0$ at generic values of the parameter $t$. The sum in this case is limited by $k=\mu$ because of vanishing the second multiplier in the nominator at larger $k$. Then, choosing in this Macdonald polynomial $t=q^{-m}$ and $\mu>2m$, we leave in the sum the terms with $0\le k\le m$ and with $\mu-m\le k\le \mu$.

However, if one chooses $t=q^{-m}$ from the very beginning keeping $\mu$ generic, the expression becomes
\be\label{2}
x_1^{\mu_1}x_2^{\mu_2}\cdot
\sum_{k=0}\left({x_2\over x_1}\right)^{k}\prod_{s=1}^{k}{\Big(1-q^{s-1-m}\Big)
\over \Big(1-q^{s}\Big)}{\Big(1-q^{\mu-s+1}\Big)
\over \Big(1-q^{\mu-s-m}\Big)}
\ee
and only the terms with $0\le k\le m$ contribute because of vanishing the first multiplier in the nominator at larger $k$. This gives rise to the BA function, which is defined at arbitrary $\mu$.

The point is that, at positive integer $\mu$, there are cancellations between the first multiplier in the nominator and the second multiplier in the denominator at $\mu-m\le k\le \mu$. Hence, the Macdonald polynomial defined only at such values of $\mu$, contains these terms, and the BA function defined at arbitrary $\mu$ does not.

Thus, we demonstrated that making a reduction of the NS function at $t=q^{-m}$ and arbitrary $\lambda$ restricts the obtained polynomial to a Weyl chamber, while making a reduction at $\vec\mu=\vec\lambda-m\vec\rho$ forming a partition gives rise to a complete symmetric polynomial.

Note that we fix a concrete Weyl chamber choosing in the NS function the ordering $i<j$ in (\ref{NS}) and the pre-factor $q^{\vec z\cdot\vec\lambda}$. This fixes the procedure (\ref{NSM}) of reduction to the Macdonald polynomial, when the symmetry between different $x_i$ is restored. At the same time, this choice in the NS function also fixes a concrete Weyl chamber associated with the BA function. One definitely can make any other choice.

In particular, this prescription of obtaining individual BA function coefficients
from corresponding NS function coefficients does solve the question that was posed
and partially answered in \cite{MMP2}: the resulting expressions are made
from rather simple factors, which are directly associated with the $A_{N-1}$ roots'
multiplicities. Note, however, that the NS solution is \textit{different} from
the one obtained in \cite{MMP2} at $N=3$: instead of depending on the
indices of non-simple roots $1$ and $3$, the extra factors in the NS ansatz depend on
seemingly random $1$ and $2$. It would be interesting, even if straightforward,
to see whether the prescription of \cite{MMP2} indeed corresponds to a different
choice of the Weyl chamber.

\section{Generalized Triads}

Now we briefly describe two available generalizations of the triad that will be discussed in detail elsewhere \cite{MMPZ}. In the Conclusion, we will discuss less developed far-going generalizations.

\subsection{Twisted Triad}

As it has been already emphasized, the triad provides eigenfunctions of the Macdonald-Ruijsenaars operator. More generally, it provides eigenfunctions of the Hamiltonians belonging to the commutative subalgebra associated with the vertical ray (0,1) of the DIM algebra \cite{MMP}. At the same time, after multiplying with a suitable exponential pre-factor, it also provides eigenfunctions of the Hamiltonians associated with the vertical ray (-1,1) \cite{MMP}. It turns out that the eigenfunctions for the rays (-1,r) are given \cite{MMP1} (see also \cite{ChF}) by the $r$-twisted BA functions $\Psi_m^{(r)}(\vec z,\vec\lambda)$ \cite{ChE}. These functions can be obtained either from linear equations, or from bilinear integrals of the usual BA functions \cite{ChE,MMP1}. One can also symmetrize these twisted BA functions to obtain {\it twisted Macdonald polynomials}. However, explicit formulas neither for the twisted BA functions, nor for the corresponding twisted NS functions are known yet \cite{MMP2}. The only known explicit formula is for $N=2$, $r=2$ \cite{MMP1}. The present status of the twisted triad is depicted in the table below.

\bigskip

\hspace{-1.4cm}
\fbox{\parbox{18.6cm}
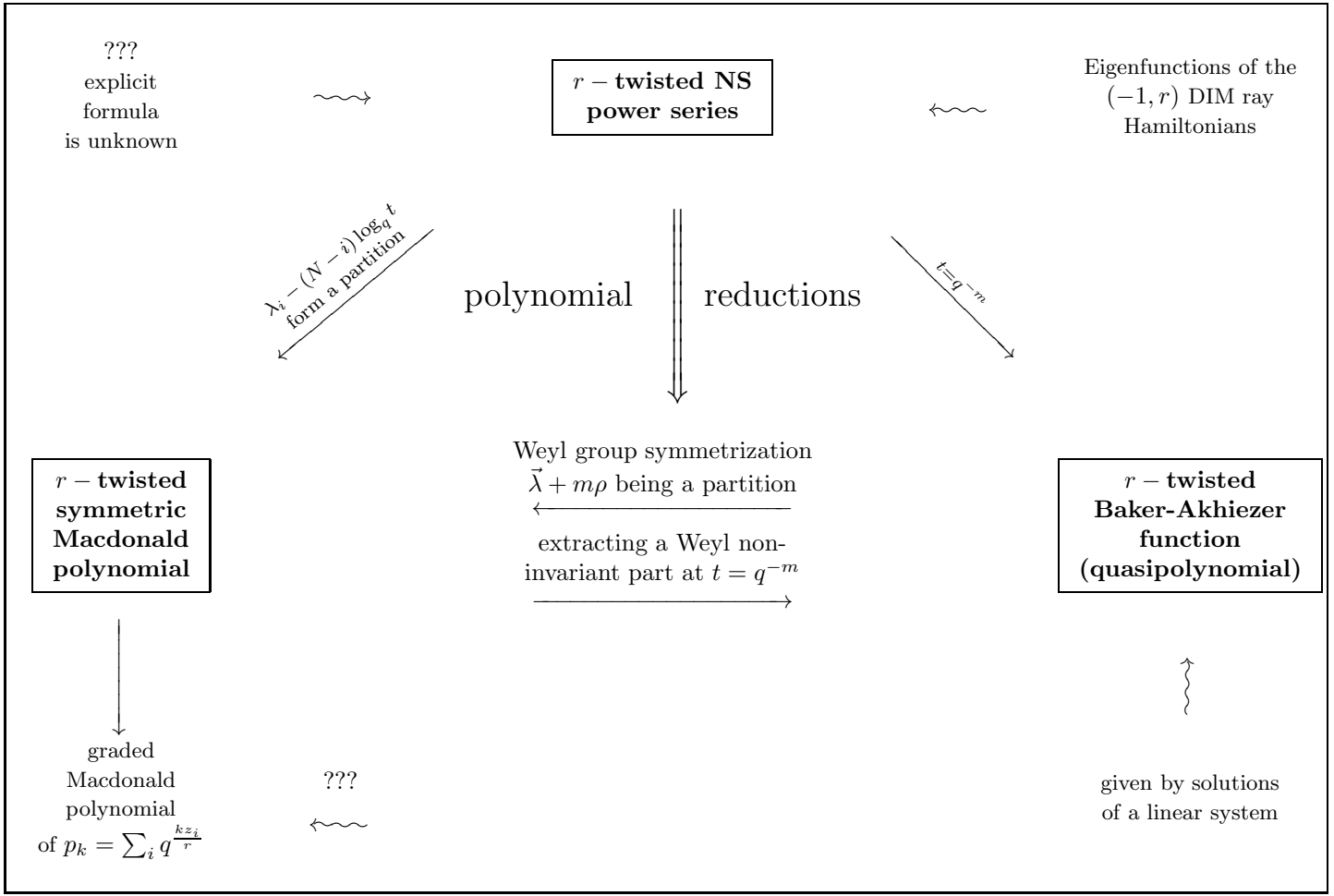
{
$$
\begin{array}{cccccr}
\begin{array}{c}???\cr
\hbox{\small explicit}\cr\hbox{\small formula}\cr\hbox{\small is unknown}\end{array}&\longsquiggly&\boxed{\begin{array}{c}
r-\hbox{\bf twisted NS}\cr
\hbox{\bf power series}
\end{array}}&{\rotatebox[origin=c]{-180}{\longsquiggly}}&\begin{array}{c}
\hbox{\small Eigenfunctions of the}\cr
(-1,r)\ \hbox{\small DIM ray}\cr
\hbox{\small Hamiltonians}
\end{array}&\\
\\
&{\rotatebox[origin=c]{40}{\(\xleftarrow{
{\footnotesize\begin{array}{c}
\lambda_i-(N-i)\log_qt\cr
\hbox{form a partition}
\end{array}}
}\)}}&\hbox{\Large polynomial {\rotatebox[origin=c]{-90}{\(\xRightarrow{{\hspace*{2.5cm}}}\)}} reductions}&{\rotatebox[origin=c]{-45}{\(\xrightarrow{\hspace*{.7cm}t=q^{-m}\hspace*{.7cm}}\)}}&\\
\\
\boxed{\begin{array}{c}r-\hbox{\bf twisted}\cr\hbox{\bf symmetric}\cr
\hbox{\bf Macdonald}\cr\hbox{\bf polynomial}\end{array}}&&\begin{array}{c}\stackrel{\begin{array}{c}\hbox{Weyl group symmetrization}\cr \vec\lambda+m\rho\hbox{ being a partition}\end{array}}{\Warrow}\cr
\stackrel{\begin{array}{c}\hbox{extracting a Weyl non-}\cr\hbox{invariant part at }t=q^{-m}\end{array}}{\bf\Earrow}
\end{array}&
&\boxed{\begin{array}{c}r-\hbox{\bf twisted}\cr\hbox{\bf Baker-Akhiezer}\cr\hbox{\bf function}\cr\hbox{\bf (quasipolynomial)}\end{array}}\\
{\rotatebox[origin=c]{-90}{\(\xrightarrow{{\hspace*{1.5cm}}}\)}}
&&&&{\rotatebox[origin=c]{90}{\longsquiggly}}\\
\begin{array}{c}
\hbox{\small graded}\cr\hbox{\small Macdonald}\cr\hbox{\small polynomial}\cr\hbox{\small of }p_k=\sum_iq^{kz_i\over r}
\end{array}
&\begin{array}{c}???\cr{\rotatebox[origin=c]{-180}{\longsquiggly}}\end{array}
&&&\begin{array}{c}\hbox{\small given by solutions}\cr\hbox{\small  of a linear system}
\end{array}
\end{array}
$$
}}

\bigskip

The question marks in the table emphasize what is unknown at the moment.

\subsection{Elliptic Triad}

Similarly to the DIM algebra, one can consider its elliptic version \cite{Saito,ell,MMZ}. Then, one can construct an elliptic version of the NS function \cite{AKMM,FOS,ELS}, which is an eigenfunction \cite{MMZ} of the degenerated Koroteev-Shakirov Hamiltonians \cite{KS} with elliptic dependence on momenta, and trigonometric dependence on coordinates. In this case, there is an explicit formula for the elliptic NS functions which additionally depends on an elliptic parameter, one can immediately make a reduction to the elliptic Macdonald polynomial \cite{AKMM,ELS} by choosing discrete set of $\lambda_i$ in the same way as it was done in the non-elliptic case, and one can also make a reduction to the elliptic BA function choosing $t=q^{-m}$ \cite{MMPZ}. However, an independent way to construct the BA function, a counterpart of the linear equations as was in the non-elliptic case is not known so far. The table that describes the present status of the elliptic triad is as follows.

\bigskip

\hspace{-1.4cm}
\fbox{\parbox{18.6cm}
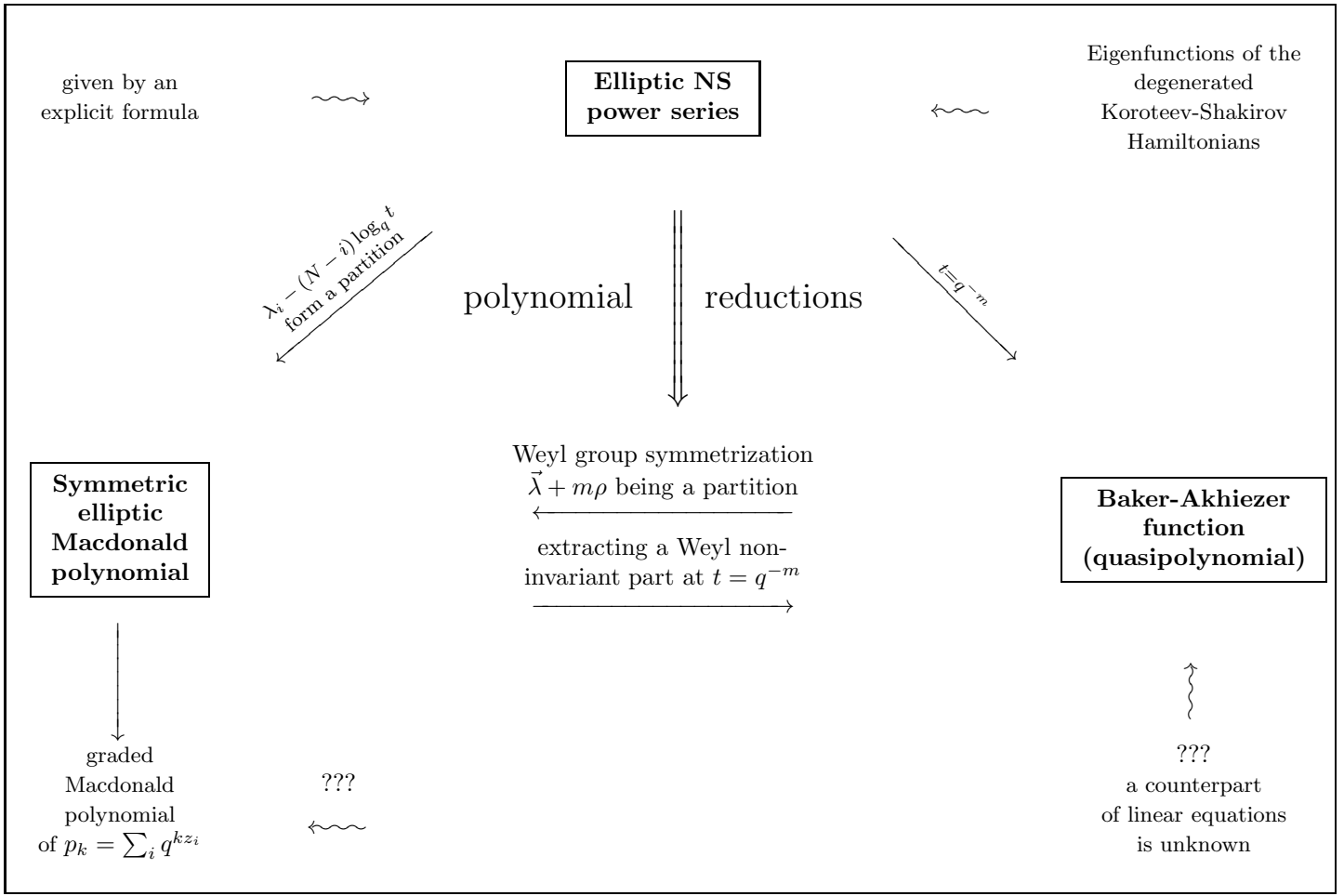
{
$$
\begin{array}{cccccr}
\begin{array}{c}\hbox{\small given by an}\cr\hbox{\small explicit formula}
\end{array}&\longsquiggly&\boxed{\begin{array}{c}
\hbox{\bf Elliptic NS}\cr
\hbox{\bf power series}
\end{array}}&{\rotatebox[origin=c]{-180}{\longsquiggly}}&\begin{array}{c}
\hbox{\small Eigenfunctions of the}\cr
\hbox{\small degenerated}\cr
\hbox{\small Koroteev-Shakirov}\cr
\hbox{\small Hamiltonians}
\end{array}&\\
\\
&{\rotatebox[origin=c]{40}{\(\xleftarrow{
{\footnotesize\begin{array}{c}
\lambda_i-(N-i)\log_qt\cr
\hbox{form a partition}
\end{array}}
}\)}}&\hbox{\Large polynomial {\rotatebox[origin=c]{-90}{\(\xRightarrow{{\hspace*{2.5cm}}}\)}} reductions}&{\rotatebox[origin=c]{-45}{\(\xrightarrow{\hspace*{.7cm}t=q^{-m}\hspace*{.7cm}}\)}}&\\
\\
\boxed{\begin{array}{c}\hbox{\bf Symmetric}\cr\hbox{\bf elliptic}\cr
\hbox{\bf Macdonald}\cr\hbox{\bf polynomial}\end{array}}&&\begin{array}{c}\stackrel{\begin{array}{c}\hbox{Weyl group symmetrization}\cr \vec\lambda+m\rho\hbox{ being a partition}\end{array}}{\Warrow}\cr
\stackrel{\begin{array}{c}\hbox{extracting a Weyl non-}\cr\hbox{invariant part at }t=q^{-m}\end{array}}{\bf\Earrow}
\end{array}&
&\boxed{\begin{array}{c}\hbox{\bf Baker-Akhiezer}\cr\hbox{\bf function}\cr\hbox{\bf (quasipolynomial)}\end{array}}\\
{\rotatebox[origin=c]{-90}{\(\xrightarrow{{\hspace*{1.5cm}}}\)}}
&&&&{\rotatebox[origin=c]{90}{\longsquiggly}}\\
\begin{array}{c}
\hbox{\small graded}\cr\hbox{\small Macdonald}\cr\hbox{\small polynomial}\cr\hbox{\small of }p_k=\sum_iq^{kz_i}
\end{array}
&\begin{array}{c}???\cr{\rotatebox[origin=c]{-180}{\longsquiggly}}\end{array}
&&&\begin{array}{c}???\cr\hbox{\small a counterpart}\cr\hbox{\small of linear equations}\cr\hbox{\small is unknown}\end{array}
\end{array}
$$
}}

\bigskip

Note that the orthogonality condition for the elliptic Macdonald polynomial as a function of the power sums, which allowed one to construct the Macdonald polynomials in the non-elliptic case, now allows one to build a set of polynomials orthogonal to the elliptic Macdonald polynomials, which are eigenfunctions of the Hamiltonians dual to the elliptic Ruijsenaars-Schneider Hamiltonians \cite{MMZ}.

\section{Conclusion}

In this note, we described the triad of the NS functions, the Macdonald polynomials and the NA functions and, briefly, its immediate generalizations to the twisted and elliptic cases. However, there are further generalizations, less developed, however, clearly understandable. Since they are not that much worked out, we do not make tables in these cases.

\paragraph{All root systems.} First of all, there is an example when both the BA function and the Macdonald polynomial are constructed, but the NS function is not. This is the case of other (than $A_{N-1}$) root systems. Indeed, we so far have been discussing only the triad associated with the $A_{N-1}$ root system. However, one can construct the Macdonald polynomials \cite{Macrs,Chrs} and the BA function \cite{Cha} for any root system. They are the eigenfunctions of the corresponding Koornwinder-Macdonald operators \cite{Macrs,Koorn}, and the BA function is again given by linear equations, straightforwardly written for a root system. However, explicit formulas for the NS functions are not known in this case.

\paragraph{Twisted elliptic Triad.} The second example is unifying the two triads of section 3: it has to be possible to construct the $r$-twisted elliptic triad, which would give rise to eigenfunctions of the elliptic DIM Hamiltonians. However, in this case, no explicit formulas for any elements of the triad are known.

\paragraph{From Shiraishi functions to the ELS functions.} The next two examples are related with lifting the NS function to the Shiraishi function \cite{Sh}
(which is a separate object from the NS function, discussed in this letter) and further to the elliptic lift of the Shiraishi (ELS) function \cite{FOS,ELS}. These ELS functions, at a particular (stationary) limit, are the eigenfunctions of the full Koroteev-Shakirov Hamiltonians \cite{MMZ,MMdell}, and they give rise to an infinite series of symmetric polynomials $M^{(k)}_\mu$ at the discrete values of $\mu$ corresponding to a partition \cite{AKMM,ELS}. However, the corresponding BA functions have not been studied yet.

\paragraph{Twisting the ELS functions.} At last, the most far-going generalization would be the ELS triad added with the $r$-twisting.
No steps in this direction were made so far.

\section*{Acknowledgements}

This work was supported by the Russian Science Foundation (Grant No.24-71-10058).

\end{document}